\documentclass[3p,twocolumn]{elsarticle}
\usepackage{graphicx}
\usepackage{algorithm, algorithmic}
\usepackage{amssymb}
\usepackage{amsmath}
\usepackage{multicol}
\usepackage{float}
\usepackage{subfigure}
\usepackage{threeparttable}
\biboptions{sort&compress}
\usepackage{url}
\usepackage{color}
\usepackage{ulem}

\begin{document}

\title{LoadCNN: A Low Training Cost Deep Learning Model for Day-Ahead Individual Residential Load Forecasting}
\author[mymainaddress,mysecondaryaddress,mythirdaddress]{Yunyou Huang}
\ead{huangyunyou@ict.ac.cn}
\author[mysecondaryaddress]{Nana Wang}
\author[mymainaddress,mythirdaddress]{Wanling Gao}
\author[mymainaddress,mythirdaddress]{Xiaoxu Guo}
\author[mymainaddress,mysecondaryaddress,mythirdaddress]{Cheng Huang}
\author[mymainaddress,mysecondaryaddress,mythirdaddress]{Tianshu Hao}

\author[mymainaddress,mysecondaryaddress,mythirdaddress]{Jianfeng Zhan\corref{mycorrespondingauthor}}
\cortext[mycorrespondingauthor]{Corresponding author}
\ead{zhanjianfeng@ict.ac.cn}

\address[mymainaddress]{State Key Laboratory of Computer Architecture, Institute of Computing Technology (ICT), Chinese Academy of Sciences (CAS), Beijing 100080, China}
\address[mysecondaryaddress]{University of Chinese Academy of Sciences, No. 19A Yuquan Road, Beijing 100049, China}
\address[mythirdaddress]{Software Systems Laboratory, ACS, ICT, CAS, Beijing 100080, China}

\begin{abstract}
Accurate day-ahead individual residential load forecasting is of great importance to various applications of smart grid on day-ahead market. Deep learning, as a powerful machine learning technology, has shown great advantages and promising application in load forecasting tasks. However, deep learning is a computationally-hungry method, and requires high costs (e.g., time, energy and $CO_2$ emission) to train a deep learning model, which aggravates the energy crisis and incurs a substantial burden to the environment. As a consequence, the deep learning methods are difficult to be popularized and applied in the real smart grid environment. In this paper, we propose a low training cost model based on convolutional neural network, namely LoadCNN, for next-day load forecasting of individual resident with reduced training cost. The experiments show that the training time of LoadCNN is only approximately $1/54$ of the one of other state-of-the-art models, and energy consumption and $CO_2$ emissions are only approximate $1/45$ of those of other state-of-the-art models based on the same indicators. Meanwhile, the prediction accuracy of our model is equal to that of current state-of-the-art models, making LoadCNN the first load forecasting model simultaneously achieving high prediction accuracy and low training costs. LoadCNN is an efficient green model that is able to be quickly, cost-effectively and environmentally-friendly deployed in a realistic smart grid environment.
\end{abstract}

\maketitle
\thispagestyle{plain}

\section{Introduction}
According to the report of National Bureau of Statistics of People's Republic of China, electricity consumption of residents was 907.16 billion kWhs in 2017~\cite{NationalData}. Residents, as important participators of smart grid, are very important to the customer-oriented applications, for example demand response (DR), demand side management (DSM), energy storage system(ESS), etc~\cite{kong2017short}. In these cases, precise day-ahead individual residential load forecasting is significant and essential to balance the generation and consumption, minimize the operating cost and decrease reserved capacity, which helps to maintain the system security and remove the requirement of expensive energy storage systems~\cite{tascikaraoglu2016short,chen2018short}.

Prediction of individual residential electricity load was firstly reported by Ghofrani et al. in 2011, and it is still a rather new area~\cite{ghofrani2011smart,kong2017short}. Only a few studies on individual residential electric load forecasting have been reported since it is an extremely challenging task. The reason lies in the huge uncertainty and volatility of electricity consumption behaviors of residents, which are difficult to be handled by traditional machine learning methods~\cite{shi2017deep,alobaidi2018robust}. Fortunately, the emergence of deep learning provides a solution to this issue, and deep learning models have shown great potential in time series prediction recently~\cite{shi2017deep,kong2017short,alobaidi2018robust,wang2019probabilistic,kong2019short,peng2019short,cai2019day}. Compared with traditional machine learning methods, deep learning shows significant superiority on individual residential electric load forecasting, owing to its powerful ability to automatically learn complex nonlinear function of relating the input x to the prediction y~\cite{zhou2017pdeep}.

However, the deep learning brings an enormous challenge: obtaining a optimal model requires high training costs (e.g., time, energy and $CO_2$ emission), which aggravates the energy crisis and puts a heavy burden on the environment. This issue has been confirmed by many recent studies, and attracted widespread attention from researchers~\cite{schwartz2019green,strubell2019energy,dalgren2019greenml,dodge2019show}. For example, Strubell et al.~\cite{strubell2019energy} investigated the training of an encoder-decoder based translation model LISA (the sequence to sequence translation task is very similar to the load forecasting task). As shown in Table~\ref{Tab1}, the total GPU time for training the model\footnote{The time is required for researching and developing the model}  was $239942$ hours, and the project spanned a period of 6 month (The model was deployed on 60 GPUs (NVIDIA Titan X (72\%) and M40 (28\%)).).  It is shown in Table~\ref{Tab1} that the estimated costs were $\$9870$ for electricity and $\$103k-\$350k$ for Google cloud computing related to researching and developing. In addition, the author pointed out that the $CO_2$ emission from training the encoder-decoder based model was five times of the $CO_2$ emission from a car within its whole lifetime~\cite{strubell2019energy}. Such a huge consumption of time, energy and the $CO_2$ emission are serious issues that cannot be ignored. 
\newcommand{\tabincell}[2]{\begin{tabular}{@{}#1@{}}#2\end{tabular}}  

\begin{table}[t]
\scriptsize
\centering
\begin{threeparttable}
\caption{The estimated costs for training an Encoder-decoder based model LISA~\cite{strubell2019energy}.\label{Tab1}}
\begin{tabular}{cccc}
\hline
&Hours&Electricity&\tabincell{c}{Cloud \\computing cost}\\
\\
\tabincell{c}{Training \\without tuning}&120&\$5&\$52-\$175\\
\\
\tabincell{c}{Training with\\ simple tuning}&2880&\$118&\$1238-\$4205\\
\\
\tabincell{c}{Training with\\ tuning}&239942&\$9870&\$103k-\$350k\\
\hline
\end{tabular}
\end{threeparttable}
\end{table}

\begin{table}[t]
\scriptsize
\centering
\begin{threeparttable}
\caption{The estimated $CO_2$ emissions~\cite{strubell2019energy}.\label{Tab2}}
\begin{tabular}{cc}
\hline
&$CO_2$ (lbs)\\
\\
Air travel, 1 passenger, $NY \leftrightarrow SF$ &1984\\
\\
Human life, avg, 1 year&11023\\
\\
Car, avg incl. fuel, 1 lifetime &126000\\
\\
\tabincell{c}{Training without tuning for \\a encoder-decoder based model} &192\\
\\
\tabincell{c}{Training with tuning for \\a encoder-decoder based model} &626155\\
\hline
\end{tabular}
\end{threeparttable}
\end{table}

In fact, there two reasons for the huge training cost of deep learning model.
On one hand, the superiority of the deep learning depends on its complex network structure ($10-1000$ millions of parameters and several to hundreds of layers) which provides powerful ability to automatically learn complex nonlinear function of relating the input $x$ to the prediction $y$~\cite{zhou2017pdeep}. However, training such a complex network even for once requires high training costs.
On the other hand, it is inevitable that developing such a complex network requires tens to thousands of experiments to adjust the structure and hyperparameters of the model for following reasons. 
\begin{itemize}
\item[a)] Develop a new deep learning model: Plenty of different network structures and hyperparameters must be considered to obtain an optimal model, which requires a number of experiments.
\item[b)] Apply latest deep learning model to individual residential electric load forecasting: Leveraging technology that people already have in their pockets for a specific task is not as simple as it appears~\cite{chen2018inferring}. The deep learning model needs a lot of adjustments to suit the individual residential electric load forecasting, which also requires a number of experiments.
What's more, due to plenty of training time cost, it is difficult to keep up with the development of deep learning technology\footnote{The update of deep learning technology is very fast as it is the most popular technology in the field of artificial intelligence~\cite{fawaz2019deep}.}.
\item[c)] Deploy a deep learning model to a specific real environment: Commonly load forecasting models are only trained on a specific small data set, which makes the model heavily depends on specific local customer behavior and local climate. It needs to adjust and retrain the model on specific data set to deploy the model in other specific environments, which also requires a number of experiments. 
\end{itemize}


However, all of the recent individual residential load forecasting researches have been focusing on promoting the forecasting accuracy and ignoring the huge training costs of deep learning model. Shi et al.~\cite{shi2017deep} firstly developed a deep learning model for individual residential load forecasting in 2017, which was an encoder-decoder based model with a novel pooling mechanism to overcome over-fitting in deep learning models. Kong et al.~\cite{kong2017short} proposed a deep learning forecasting framework based on long short-term memory (LSTM), aiming to address the volatility of the electricity consumption behaviors of individual residents. Wang et al~\cite{wang2018short} developed a gated recurrent unit (GRU) model which was a popular recurrent neural network (RNN) model to forecast the next-day load of individual resident. Wang et al.~\cite{wang2019probabilistic} reported an LSTM model with a new loss function (pinball loss), week information and hour information to forecast the load of the individual resident. Recently, Kong et al.~\cite{kong2019short} proposed an LSTM-based framework to handle the high volatility and uncertainty of individual residential load. Though RNN-based model is the mainstream model for sequence prediction, and it achieves state-of-the-art performance in individual residential load forecasting tasks~\cite{wang2019probabilistic,kong2019short}, the training of RNN-based model is time-consuming and requires a large amount of computing resources since the complex mechanism of RNN. In addition, the RNN-based model is difficult to parallel which requires more training time.

According to the reasons for huge training cost which mention above, two ways are able to reduce the training costs: 1)reduce training cost of each training. 2)reduce the total number of training. In this paper, we propose a convolutional neural network (CNN) based model LoadCNN to reduce the training cost of each training and simultaneously keep an excellent forecasting accuracy. The experiments show that training current state-of-the-art models requires enormous cost, and our model significantly outperforms current state-of-the-art methods. Meanwhile, our model is able to reach the standard of the state-of-the-art performance in forecasting accuracy.
\begin{figure*}
\centering
\includegraphics[width=6.5in, height=2.1315in]{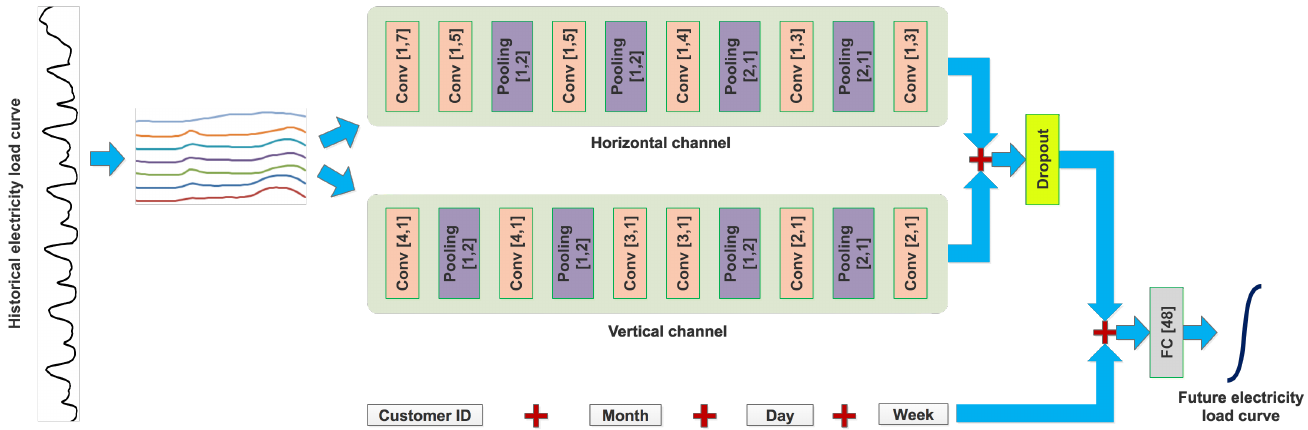}
\caption{The structure of LoadCNN.~\label{Fig1}}
\end{figure*}

The contributions of this paper are as following four aspects:
\begin{itemize}
\item[a)] New problem: Training costs are firstly considered in load forecasting task, which are important issues that have been ignored in previous researches.
\item[b)] New view: Affected by the life and work habits of residents, many regularities will exist in a day and in the same time periods of different days. In order to make it easier for CNN to obtain the regularity of electricity consumption in the same time periods of different days, we convert the electricity load curve into two-dimension so that electricity consumption values of the same time period are nearby to each other.
\item[c)] New model: We propose a novel load forecasting model LoadCNN based on CNN. Our model contains two channels to directly extract the consumption regularity among a day and the consumption regularity among a week at the same time period respectively. Besides, LoadCNN only contains $7$ layers and all of the convolution layers are one-dimension, which effectively reduces the training costs of our model.
\item[d)] The training time of our model is only approximate $1/54$ of other state-of-the-art models, the energy consumption and $CO_2$ emissions are only approximate $1/45$ of other state-of-the-art models. Meanwhile, it equals to the current state-of-the-art performance in terms of prediction accuracy.
\end{itemize}

The rest of this paper is structured as follows. 
Section~2 introduces our innovative approach. Section~3 describes the experiment deployment.
Section~4 presents and discusses the results. Section~5 draws a concluding remark.
\section{Methodology}
 In this section, we give a formal definition of day-ahead forecasting and propose a novelty CNN-based model LoadCNN for day-ahead load forecasting.

\subsection{Day-ahead Individual Residential Load Forecasting}

Load curve represents electricity consumption behaviors of individual residents, which is very important to various customer-oriented applications in smart grid. Load curve of an individual resident is denoted as $X=\{x_t \ | \ t>0\  and\  t\in \mathbb{R}\}$ in this paper. Also, we use the historical load curve of the individual resident $X_h=\{x_t \ | \ t\ge T-s\  ,\  t< T\  and\ s>0\}$ to predict future load curve of individual resident $X_f=\{x_t \ | \ t\ge T\  ,\  t< T+l\  and\ l>0\}$ . Here, $x$ is the electricity consumption value in the load curve $X$, $s$ is the length of the historical load curve $X_h$, $l$ is the length of future load curve $X_f$, and time step $T$ divides load curve into input and output of load forecasting task.

In this work, we focus on day-ahead load forecasting based on historical load curve of past $7$ days, with $l=48$ and $s=336$ (half an hour interval, 48 data points for a day). Predicted load $\hat{X_f}=\{\hat{x_t} \ | \ t\ge T\  ,\  t< T+48\}$ is defined by Equation~\ref{eq1}.
\begin{equation}\label{eq1}
  \hat{X_f}=f(X_h), \quad X_h=\{x_t \ | \ t\ge T-336\  ,\  t< T\}
\end{equation}
The object of day-ahead load forecasting task is to minimize prediction $loss$ defined by Equation~\ref{eq2}.
\begin{equation}\label{eq2}
loss(\hat{X_f},X_f)\\
  =\sqrt{\frac{1}{48}\sum_{i=T}^{T+48}(x_i-\hat{x_i})^2}
\end{equation}

\subsection{Details of LoadCNN}
In this section, we will elaborate our proposed method LoadCNN. In addition, we will also introduce day-ahead individual residential load forecasting algorithm which is based on LoadCNN. 

\subsubsection{Data preparation}

Data preprocessing is an essential step of the load forecasting. In our paper, five types of data are fed into LoadCNN, including historical load curve $\textbf{L}$, individual residential $\textbf{ID}$, month $\textbf{M}$, day $\textbf{D}$, and week $\textbf{W}$. The details of them are explained as follows:
\begin{itemize}
\item[a)] Historical load curve $\textbf{L}$ is a sequence of energy consumptions of the past $7$ days, and the size of $\textbf{L}$ is $336$. In order to easily extract the relationship among a day and the relationship between different days by CNN, we reshape $\textbf{L}$ into a matrix with the shape [7,48].
\item[b)] The customer ID of individual residential $\textbf{ID}$ is several vectors that are encoded by one hot encoder. Since the number of customer $N$ is generally large ($N=929$), we utilize two vectors to uniquely represent a customer to obtain the vector with a relative smaller size. The size of each vector is $31=\ulcorner\sqrt[2]{N}\urcorner$, and the size of $\textbf{ID}$ is $62=2\times31$. Similarly, if necessary, we can use $k$ vectors to represent $\textbf{ID}$.
\item[c)] External factors such as lifestyle of customer and weather will have a direct impact on electricity consumption behavior. We consider the month, date, and week information in our model as the following folds: 1) The month index provides some information regarding weather (such as summer is hot). 2) The week index provides some information regarding work (such as weekend is holiday). 3) The date index provides some information regarding special behavior (such as shopping after pay). The month $\textbf{M}$, day $\textbf{D}$ and week $\textbf{W}$ are encoded by one hot encoder and belong to load curve to predict. The sizes of $\textbf{M}$, $\textbf{D}$ and $\textbf{W}$ respectively are $12$, $31$ and $7$.
\end{itemize}

\subsubsection{Load forecasting model}
Compared with RNN, CNN has a simpler neural network structure and achieves state-of-the-art performance in image processing realm~\cite{voulodimos2018deep}. Thus, we seek to develop an energy-saving and efficient green model based on CNN. In this paper, we proposed a CNN-based model named LoadCNN for load forecasting which is shown in Figure~\ref{Fig1}. The main components of the LoadCNN are horizontal channel and vertical channel.
\begin{itemize}
\item[a)] Horizontal channel: This channel is used to extract the electricity consumption regularity among a day. It contains $6$ convolution layers shaped like \textbf{[1,N]}, and $4$ maxpooling layers. The convolution layers with $16$, $24$, $24$, $64$, $64$, and $64$ kernels, and all of the convolution layers are activated by the Rectified Linear Unit (ReLU) function.
\item[b)] Vertical channel: This channel is used to extract the electricity consumption regularity between different days. It contains $6$ convolution layers shaped like \textbf{[N,1]}, and $4$ maxpooling layers. The convolution layers also with $16$, $24$, $24$, $64$, $64$, and $64$ kernels, and all of the convolution layers are activated by the Rectified Linear Unit (ReLU) function.
\end{itemize}
Besides, the \textbf{ID, M, D, W} and the data from the channels will be contacted and fed into a full connected layer.


Unlike the previous load forecasting works which only simply adjust existing models, we designed LoadCNN according to the electricity consumption habits of residents to extract the electricity consumption regularity and reduce the training costs. The LoadCNN consists of two channels, which respectively extract the consumption regularity among a day and the consumption regularity among a week in the same time period. Thus LoadCNN is able to easily extract the electricity consumption patterns of the consumer. In addition, the one-dimension convolution layers make the structure of LoadCNN more simple and easy to be trained. For example, a two-dimension convolution layers with the shape \textbf{[4,7]}, is able to simultaneously capture the electricity consumption regularity in horizontal and vertical, and to replace the first layer of LoadCNN.  However, the two-dimension convolutional kernel contains $28=4\times7$ elements, much more than the elements of the two one-dimension convolutional kernels in the first layer of LoadCNN, which has $11=4\times1+1\times7$ elements. It brings more computation, and raises the training cost of the model.

\begin{algorithm}
    \renewcommand{\algorithmicrequire}{\textbf{Input:}}
    \renewcommand{\algorithmicensure}{\textbf{Output:}}
    \caption{The algorithm for individual residential electric load forecasting}
    \label{alg1}
    \begin{algorithmic}[1]
    \REQUIRE Load dataset $\Psi$ of residents demand from smart meters.
    \ENSURE The predicted Load of individual residents $\hat{X_f}$ and the root mean squared error (RMSE), normalised root mean squared error (NRMSE), and mean absolute error(MAE).
    \STATE Clean and pre-process the load data and obtain a dataset $\Psi_0=\{X_h,X_f\}$. $X_h$ is the historical load and $X_f$ is the target.
    \STATE Divide $\Psi_0$ into training set $\Psi_{tr}$, validation set $\Psi_{va}$, and test set $\Psi_{ts}$.
    \STATE Initialize all learnable parameters $\theta_0$ in LoadCNN.
    \STATE The best parameters $\theta_{best}=\theta_0$.
    \STATE The best validation loss $loss_{best}=0.55$.
    \FOR {Current epoch $k^{th} < $ Max epoch}
    \WHILE{ Any instances in $\Psi_{tr}$ not are selected in this epoch.}
    \STATE Select a batch of instances $\Psi_b$ from $\Psi_{tr}$.
    \STATE Find $\theta_i$ by minimizing the $loss$ defined by Equation~\ref{eq2} with $\Psi_b$.
    \IF {$k^{th}\ \% \ 100==0$}
    \STATE Randomly select a batch of instances $\Psi_{b\_va}$ from $\Psi_{tr}$.
    \STATE Calculate $loss_{va}$ defined by Equation~\ref{eq2} with $\Psi_{b\_va}$.
    \IF {$loss_{best}<loss_{va}$}
    \STATE $loss_{best}=loss_{va}$
    \STATE $\theta_{best}=\theta_i$
     \ENDIF
    \ENDIF
    \ENDWHILE
    \ENDFOR
    \STATE Forecast the $\hat{X_f}$ by LoadCNN with $X_h \in \Psi_{ts}$.
     \STATE Calculate the RMSE, NRMSE, and MAE with the $\hat{X_f}$  and $X_f \in \Psi_{ts}$. 
    \end{algorithmic}
\end{algorithm}
\subsubsection{Algorithm}
The algorithm designed includes three parts as shown in Algorithm~\ref{alg1}: 1) data pre-processing, 2) network training, and 3) evaluation.

\section{Experiment Deployment}
\subsection{Data description}
To evaluate the performance of LoadCNN, we conduct the experiments on a large-scale smart meter dataset from Smart Metering Electricity Customer Behaviour Trials (CBTs) in Ireland~\cite{smart2011metering}.
The data is collected from over $5000$ Irish customers for the period of $536$ days between 1st July 2009 and 31st December 2010. The smart meter data is half-hourly sampled electricity consumption (kWh) data from each customer.

In CBTs, we selected the customers which meet the condition that residential customers with the controlled stimulus and controlled tariff because of the following two aspects:
(1) selected customers were billed on existing flat rate without any DSM stimuli. (2) selected customers are the most representative\footnote{The majorities of consumers outside trial are of the type~\cite{shi2017deep}}.
Finally, $929$ residential customers are selected to verify our method.

To verify our method, we divide the dataset into three sets: training set, validation set, and test set. The test set contains all the data of the last $30$ days. The validation set contains data of $60$ days which are randomly selected from the $8th-506th$ days. The training set contains all of the rest data.

\subsection{Experiment Setup}

All of models for all customers are built on a server with two Intel Xeon E5-2630 v4 processors, $256$ GB of memory and four NVIDIA Titan Xp GPUs. Server system is Linux 3.10.0-327.el7.x86\_64. In addition, all of the models are implemented by the TensorFlow-gpu 1.10.0v library~\cite{tensorflow} and Python 3.6.7v.

The parameters for all models are presented as follows: batch size=$64$, max epoch=$65$, hidden neuron number of RNN=$128$, learning rate=$0.0015$, decay rate=$0.96$. All of the RNN-based model are adopted dropout technology, and dropout rate=$0.5$. In addition, in order to facilitate the comparison of training time and energy consumption, each model runs on only one GPU.

\subsection{Metrics}
In this work, three widely used metrics are applied to evaluate the accuracy of LoadCNN: root mean squared error (RMSE), normalised root mean squared error (NRMSE), and mean absolute error (MAE).
\begin{equation}\label{eq3}
RMSE=\sqrt{\frac{\sum_{i=1}^{N}(x_i-\hat{x_i})^2}{N}}
\end{equation}
\begin{equation}\label{eq4}
NRMSE=\frac{RMSE}{x_{max}-\hat{x_{min}}}
\end{equation}
\begin{equation}\label{eq5}
MAE=\frac{\sum_{i=1}^{N}|(x_i-\hat{x_i})|}{N}
\end{equation}
Here, $\hat{x_i}$ is the predicted value, $x_i$ is the actual value, $x_{max}$ and $x_{min}$ are the maximum and minimum value of $x_i$ respectively. $N$ is the number of point $x_i$ in the test set.

\begin{figure}
\centering
\includegraphics[width=3in, height=1.811in]{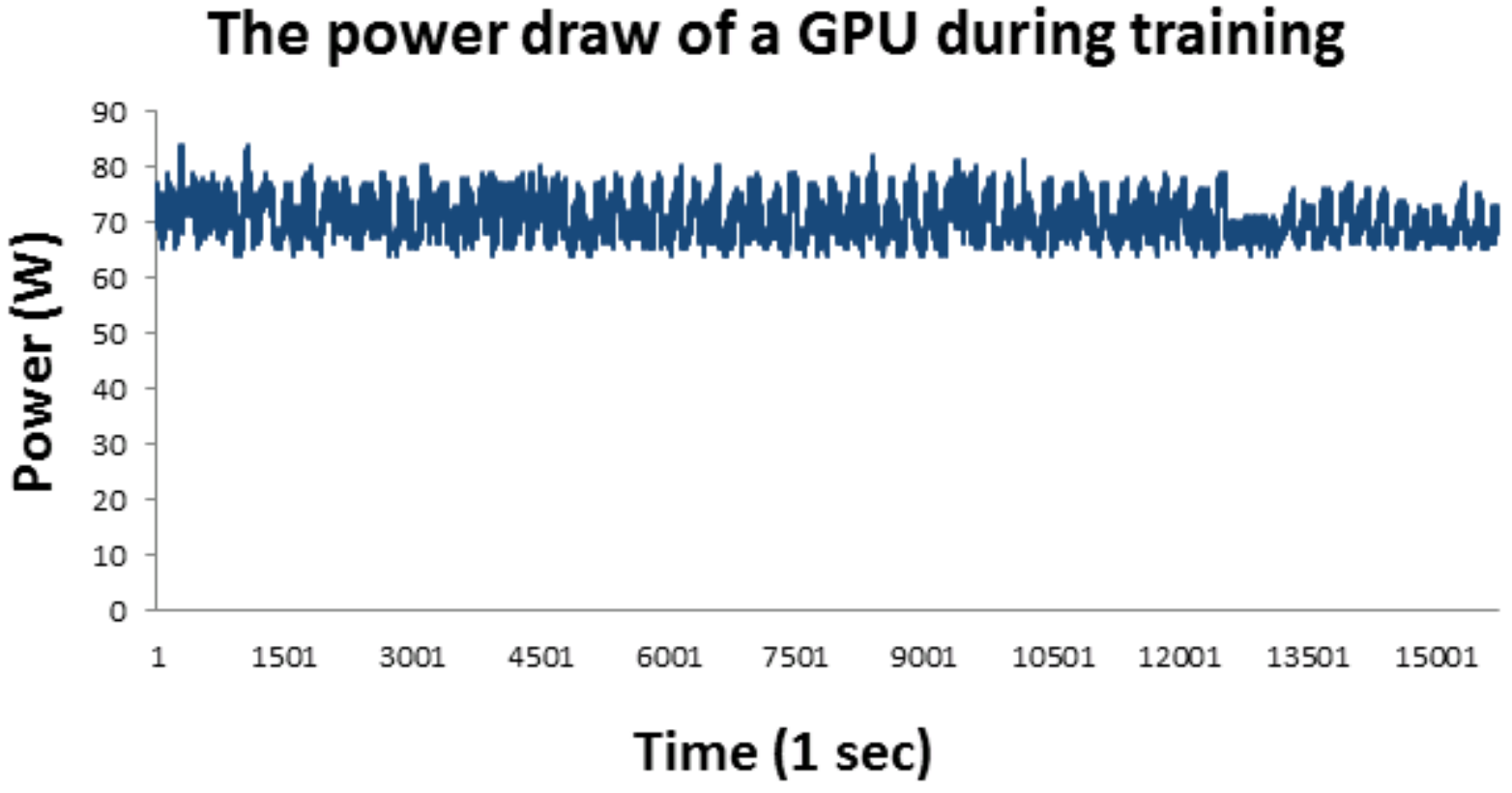}
\caption{The power draw of a GPU during training a encoder-decoder model.~\label{Fig2}}
\end{figure}

Meanwhile, energy efficiency and training efficiency are also needed to measure in our work.
Energy consumption ($EC$) is defined in Equation~\ref{eq6} as GPU consumes the most part of energy.
\begin{equation}\label{eq6}
EC=\frac{P\times TT\times PUE\times NT}{1000}
\end{equation}
Here, $P$ is the power of GPU during training the model. The $TT$ represents the training time of a model for one training. $PUE$ is the power usage effectiveness and accounts for the additional energy that is required to support the compute infrastructure (mainly cooling)~\cite{strubell2019energy}.
$NT$ is the number of times to train a model.
The detailed settings of the parameters are as follows.
\begin{itemize}
\item[a] $P$: as Figure~\ref{Fig2} shows, the differences of the power drawn of a GPU during training a model are negligible. Thus, to simplify the problem and minimize the impact of monitoring procedures on training, we randomly select the average power within 30 minutes during model training as $P$ for model training.
\item[b] $PUE$: its coefficient is set as 1.58 (global average for data center) according to the study~\cite{strubell2019energy}. 
\item[c] $NT$: 
In general, hyperparameter tuning is a big topic and essential to obtain the best forecasting performance~\cite{kong2019short}. In the recent~\cite{strubell2018linguistically} work, to obtain the best performance of an encoder-decoder model, the author did 4789 trials~\cite{strubell2019energy}. The task of the model, which is a task forecasting of sequence to sequence, is similar to the day-ahead individual resident load forecasting.
Thus, In our paper, to simplify the problem we assume that NT=$1000$ trials are required to obtain the best performance of a model.
\end{itemize}
 The reasons for assumptions above are as follows: 1) Every model runs on the same sever.  2) Every model runs on a NVIDIA Titan Xp GPU only. 3) Most of the energy consumption of training a model is on GPU.

The $CO_2$ emissions $CO_2e$ is estimated by Equation~\ref{eq7} according to U.S. Environmental Protection Agency~\cite{strubell2019energy}.

\begin{equation}\label{eq7}
CO_2e=0.954EC
\end{equation} 

\subsection{Compared Models Setup}
We use $11$ models from four types of popular deep learning methods as benchmarks in present work: classic RNN-based model, RNN and CNN-based model, encoder-decoder-based models and CNN-based model.
\begin{itemize}
\item LSTM, a most popular RNN model for time series prediction, is commonly used for load forecasting since 2017~\cite{kong2017short}. In our paper, we transform the model into a day-ahead load forecasting model to compare with our model.
\item LSTM-Week is a load forecasting model proposed in 2019. It uses a new loss function and considers the week and hour information~\cite{wang2019probabilistic}. 
\item LSTM-EID is also a load forecasting model proposed in 2019. It considers the week, record point position in a day and holiday information~\cite{kong2019short}. In order to compare with our model, we also transform the model into a day-ahead load forecasting model.
\item GRU is another popular RNN model proposed in 2018 for time series prediction and applied to day-ahead load forecasting~\cite{wang2018short}. The model considers date, weather and temperature information. 
\item Skip-RNN is an RNN model that is able to capture long term dependencies and
relieve vanishing gradients when the model is trained on long sequences~\cite{campos2018skip}. Since the length of input in this work is $336=7*48$, the Skip-RNN is considered as a benchmark.
\item LSTM-CNN is a model that mixes typical LSTM and CNN which is similar to a famous model--inception models~\cite{szegedy2017inception}. The types of LSTM-CNN model have been used to load forecasting on area level and  industrial distribution complexes~\cite{tian2018deep,kim2019recurrent}. In order to compare with our model, we transform LSTM-CNN model into a day-ahead load forecasting model on individual resident level.
\item seq2seq is a LSTM-based encoder-decoder model which is the most popular model for the forecasting of sequence to sequence.
\item seq2seq-pooling was proposed in 2017 to relieve the overfitting in load forecasting~\cite{shi2017deep}. In order to compare with our model, we transform the model into a day-ahead load forecasting model and use the dropout technology to further relieve the overfitting.
\item seq2seq-attention is an encoder-decoder model that combines the attention mechanism to handle the long sequences~\cite{bahdanau2014neural}.
\item Temporal convolutional network (TCN) is recently proposed to handle sequence and
achieves the state-of-the-art performance in many sequence modeling tasks~\cite{bai2018empirical}. It has been used to load forecasting on individual resident level~\cite{voss2018residential}. In order to compare with our model, we also transform the model into a day-ahead load forecasting model.
\item ResNet, a CNN-based model, is the state-of-the-art method of image recognition tasks~\cite{he2016deep}.
\end{itemize}

\section{Results and discussion}
In this section, we present and discuss the results of the experiments in terms of training efficiency, energy consumption, environmental costs and prediction accuracy. In addition, we also investigate the effect of number of layers in deep learning model since the deeper the model, the more complex the network structure of the model, and the more training time is needed which results in more energy consumed and $CO_2$ emitted.

\begin{table*}[t]
\tiny
\centering
\begin{threeparttable}
\caption{Performance comparison.\label{Tab3}}
\begin{tabular}{ccccccccccc}
\hline
\tabincell{c}{Network\\ Architecture}&$\tabincell{c}{TT \\without \\tuning (h)}$&$\tabincell{c}{Power (W)}$&$\tabincell{c}{EC (kWh)}$&$\tabincell{c}{$CO_2e (lbs)$}$&$\tabincell{c}{Easy to\\ parallel}$&$\tabincell{c}{RMSE\\(kWh)}$&$\tabincell{c}{NRMSE\\(kWh)}$&$\tabincell{c}{MAE\\(kWh)}$&$Layers$&$Steps$\\
\\
$LSTM/3$&164.42&66.1656&17188.7378&16398.0559&No&0.6192&0.0473&0.3636&3&336\\
\\
$LSTM/5$&239.22&-&-&-&No&0.6157&0.0470&0.3511&5&336\\
\\
$LSTM/8$&365.65&-&-&-&No&0.7375&0.0563&0.4085&8&336\\
\\
$LSTM-Week$&164.73&68.5650&17845.6456&17024.7459&No&0.6246&0.0477&0.3665&3&336\\
\\
$LSTM-EID$&161.58&68.3967&17461.4313&16658.2055&No&0.6153&0.0470&0.3639&3&336\\
\\
$GRU$&170.30&64.5683&17373.6508&16574.4629&No&0.6156&0.0470&0.3487&3&336\\
\\
$Skip-RNN$&190.33&\textbf{64.1756}&19298.9762&18411.2233&No&\textbf{0.6147}&\textbf{0.0469}&0.3477&3&336\\
\\
\\
$LSTM-CNN$&153.2&67.3422&16300.5835&15550.7567&No&0.6184&0.0472&0.3583&3-8&336-1\\
\\
\\
$seq2seq/3$&165.02&72.4456&18888.8572&18019.9698&No&0.6641&0.0507&0.4101&3-3&336-48\\
\\
$seq2seq/5$&274.28&-&-&-&No&0.6771&0.0517&0.4806&5-5&336-48\\
\\
$seq2seq/8$&389.95&-&-&-&No&0.6881&0.0525&0.4941&8-8&336-48\\
\\
$seq2seq-pooling/3$&164.12&66.4822&17239.4727&16446.4570&No&0.6713&0.0513&0.3922&3-3&336-48\\
\\
$seq2seq-pooling/5$&246.22&-&-&-&No&0.6581&0.0503&0.4332&5-5&336-48\\
\\
$seq2seq-pooling/8$&382.95&-&-&-&No&0.7252&0.0554&0.5474&8-8&336-48\\
\\
$seq2seq-attention$&180.33&87.1394&24827.8798&23685.7973&No&0.6549&0.0500&0.4005&3-3&336-48\\
\\
\\
$TCN$&20.55&218.3589&7089.8951&6763.7599&\textbf{Yes}&0.8770&0.0670&0.4731&8&1\\
\\
$ResNet$&7.15&187.5428&2118.6710&2021.2121&\textbf{Yes}&0.6261&0.0478&0.3673&34&1\\
\\
$\textbf{LoadCNN (Our)}$&\textbf{2.85}&80.2228&\textbf{361.2433}&\textbf{344.6261}&\textbf{Yes}&0.6152&0.0470&\textbf{0.3469}&7&1\\
\hline
\end{tabular}
\end{threeparttable}
\end{table*}

\subsection{Training efficiency, energy consumption and environmental costs} 
As shown in Table~\ref{Tab3}, our model not only almost achieves the second highest prediction accuracy, which is only 0.0005 behind the first place, but also obtains superior performance in training time, energy consumption and $CO_2$ emissions compared with all the other models.
Specifically, in training time,
LoadCNN takes the shortest time that is only approximate $1/54$ of other RNN-based models. 
What's more, LoadCNN is based on CNN and very easy to parallel. 
Thus, the training time of LoadCNN is able to be further reduced by simply adjusting the code of implementation and adding more GPUs.
As for energy consumption and $CO_2$ emissions, LoadCNN is also only approximate $1/45$ of other RNN-based models.

The reason for the results is that LoadCNN has a simple network structure which is easy to be trained. However, contrary to our model: 1) The step of RNN-based model is $336$ which is a quite large number and makes the structure of the model very complex when training, though the layer of the RNN based model is $3$. 2) The other CNN based models are also much more complex than our model because they have more layers.

In addition, compared with the experiment, the training time, energy consumption and $CO_2$ emissions of the model will be more in reality. In this experiment the training set only contains the data from $929$ customers for $439=536-30-60-7$ days. In the optimal case, the energy consumption and training time of the mainstream models are expected to exceed $16000$ kWhs and $150000=150 \times1000$ h ($17$ years) respectively (Of course, if there are enough GPUs, we can perform multiple parameter adjustment experiments at the same time to reduce the training time.). However, in real
environment the training set should contain hundreds of thousands or even more customers, which will significantly increase the time and energy consumption of the training model. 
Therefore, training efficient and low-energy models like our model is significant.

\subsection{The prediction accuracy of different deep learning models.}
\begin{figure}
\centering
\includegraphics[width=3in, height=1.811in]{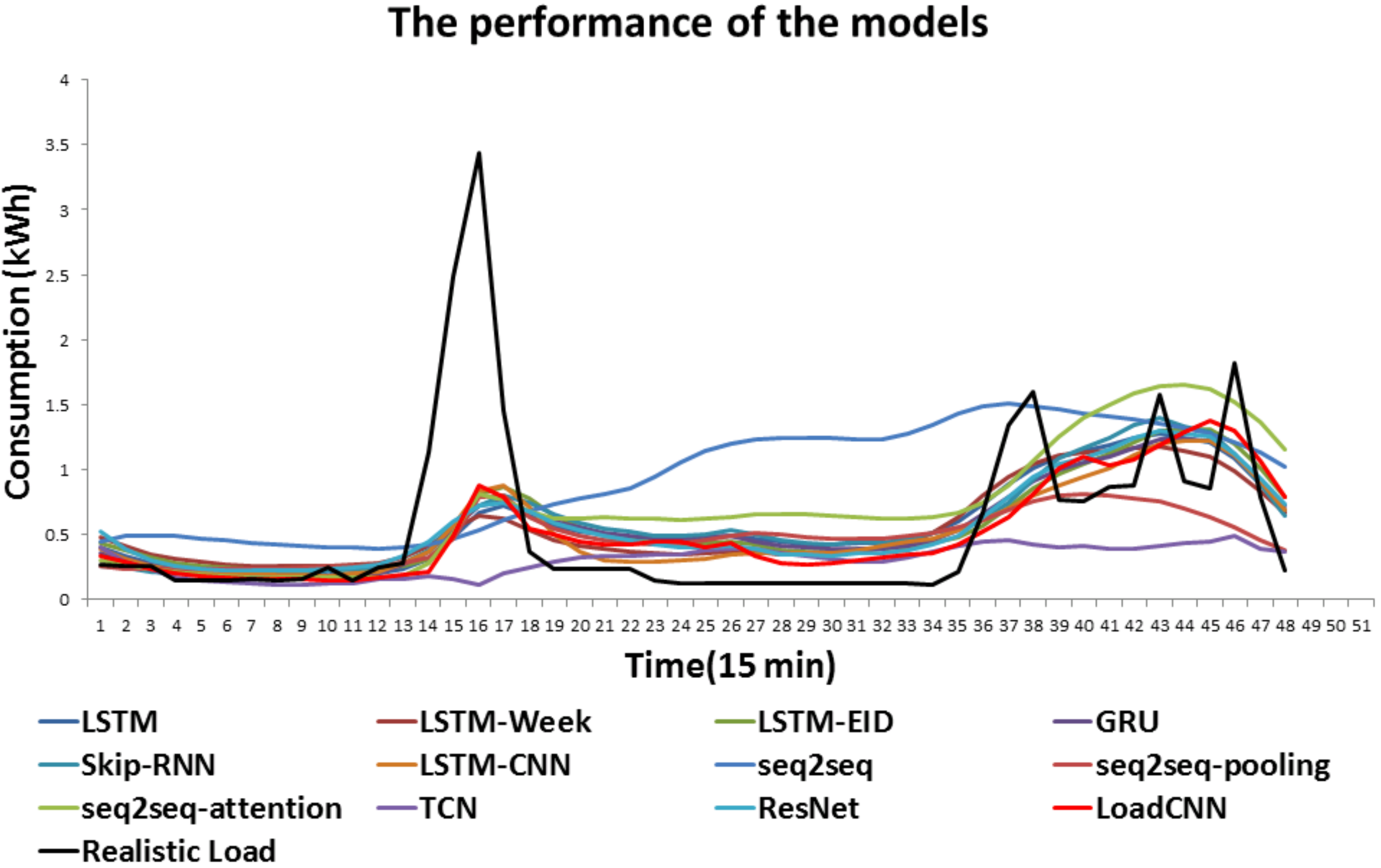}
\caption{The forecasting loads by different models and the realistic load.~\label{Fig3}}
\end{figure}

It is found that the prediction accuracy is hard to improve only by constructing different deep learning models.
Specially,
as shown in Table~\ref{Tab3}, 
the best performance ones of classical RNN-based models, RNN-CNN-based models, and CNN-based models have little difference in the accuracy metrics $RMSE$, $NRMSE$, and $MAE$. Usually, the tiny difference is likely to be eliminated by adjusting hyperparameters. 
Consistent with Table~\ref{Tab3}, as shown in Figure~\ref{Fig3} except for encoder-decoder based models and TCN model, the prediction performances of other models are not much different. 
It means that it is difficult to use current deep learning technology to make a major breakthrough in forecasting accuracy of day-ahead individual residential load forecasting. 

In order to improve the accuracy of the forecast, we need to pay more attention to obtain the information about personal activities and external information, as the electricity consumption of the household is extremely dependent on the randomness of individual human behaviors and external factors\footnote{For example, both the  business trip of the residents and the change of indoor temperature will cause electricity consumption changes which are difficult to predict only by the historical load curve of the individual resident.}. 

In addition, it is worth noting that there is a large gap between the performance of the encoder-decoder based models and state-of-the-art models.
This can be explained by the mechanism of the decoder. In the day-ahead individual resident load forecasting task, the decoder predicts the value of the current point based on the value of the previous point and the current state of model. Unfortunately, we can not directly obtain the value of the previous point, and the predicted value of the previous point is used to replace the actual value. Therefore, the forecasting errors will be accumulated and amplified.

Finally, it is also worth noting that compared with the previous very short-term (such as 15-min-ahead) load forecasting work which predicted the electricity consumption more accurately, the day-ahead load forecasting tends to predict electricity consumption pattern of the customers. For example, as shown in Figure~\ref{Fig3}, the early peak and the three later peaks of the actual load curve cannot be accurately predicted.

\subsection{Effect by the number of layer in model}
The recent revival of neural networks has benefited from the development of computer hardware that has made neural networks deeper and deeper which is the main cause of the high complexity of deep learning model.

In general, the deeper the neural network is the more precise the prediction is. However, as shown in the Table~\ref{Tab3}, deeper models of LSTM, seq2seq, seq2seq-pooling, CNN do not perform better than shallow models. What's more, $8$-layer LSTM, seq2seq and seq2seq-pooling models have terrible performance. It means that on the one hand, the most powerful means of deep learning -- increasing depth can no longer help improve the accuracy of the model. On the other hand we need to develop new technologies to solve the over-fitting problem.

As a conclusion, it is unnecessary to make the model deeper, which leads to a more complex network structure that will translate into more training time, more energy consumption and more $CO_2$ emission.


\section{Conclusions}
Day-ahead individual residential load forecasting is very important to real applications (such as demand response) of smart grid. Deep learning models have became commonly used methods in load forecasting. However, the deep learning models are computationally-hungry, which requires plenty of training costs. All of the previous load  forecasting works have been only focusing on improving prediction accuracy and ignore training costs.

To save resources and promote the application of deep learning models,
we propose and develop an efficient green CNN-based model LoadCNN. It not only equal to the current state-of-the-art performance but also has huge advantages in training efficiency, energy consumption and environmental costs.
The experimental results on public and large-scale dataset show that the training time of our model is only approximate $1/54$ of other state-of-the-art models, the energy consumption and $CO_2$ emissions are only approximate $1/45$ of other state-of-the-art models. 
In addition, it is found that it is difficult to improve the accuracy by simply adjusting the hyperparameters or structure of deep learning models.
In the future, to improve the accuracy, we must obtain more relevant information (such as human activities). 

\section*{Acknowledgements}
We are very grateful to CER Smart Metering Project - Electricity Customer Behaviour Trial, 2009-2010 and ISSDA. This work is supported by the Major Program of National Natural Science Foundation of China (Grant No. 61432006).

\section*{Competing interests} The authors declare  no competing interests.


\begin{thebibliography}{10}
\expandafter\ifx\csname url\endcsname\relax
  \def\url#1{\texttt{#1}}\fi
\expandafter\ifx\csname urlprefix\endcsname\relax\def\urlprefix{URL }\fi
\expandafter\ifx\csname href\endcsname\relax
  \def\href#1#2{#2} \def\path#1{#1}\fi

\bibitem{NationalData}
N.~B. of~Statistics~of China, Annual data,
  \url{http://data.stats.gov.cn/easyquery.htm?cn=C01}, accessed June 12, 2019.

\bibitem{kong2017short}
W.~Kong, Z.~Y. Dong, D.~J. Hill, F.~Luo, Y.~Xu, Short-term residential load
  forecasting based on resident behaviour learning, IEEE Transactions on Power
  Systems 33~(1) (2017) 1087--1088.

\bibitem{tascikaraoglu2016short}
A.~Tascikaraoglu, B.~M. Sanandaji, Short-term residential electric load
  forecasting: A compressive spatio-temporal approach, Energy and Buildings 111
  (2016) 380--392.

\bibitem{chen2018short}
K.~Chen, K.~Chen, Q.~Wang, Z.~He, J.~Hu, J.~He, Short-term load forecasting
  with deep residual networks, IEEE Transactions on Smart Grid.

\bibitem{ghofrani2011smart}
M.~Ghofrani, M.~Hassanzadeh, M.~Etezadi-Amoli, M.~S. Fadali, Smart meter based
  short-term load forecasting for residential customers, in: 2011 North
  American Power Symposium, IEEE, 2011, pp. 1--5.

\bibitem{shi2017deep}
H.~Shi, M.~Xu, R.~Li, Deep learning for household load forecasting—a novel
  pooling deep rnn, IEEE Transactions on Smart Grid 9~(5) (2017) 5271--5280.

\bibitem{alobaidi2018robust}
M.~H. Alobaidi, F.~Chebana, M.~A. Meguid, Robust ensemble learning framework
  for day-ahead forecasting of household based energy consumption, Applied
  energy 212 (2018) 997--1012.

\bibitem{wang2019probabilistic}
Y.~Wang, D.~Gan, M.~Sun, N.~Zhang, Z.~Lu, C.~Kang, Probabilistic individual
  load forecasting using pinball loss guided lstm, Applied Energy 235 (2019)
  10--20.

\bibitem{kong2019short}
W.~Kong, Z.~Y. Dong, Y.~Jia, D.~J. Hill, Y.~Xu, Y.~Zhang, Short-term
  residential load forecasting based on lstm recurrent neural network, IEEE
  Transactions on Smart Grid 10~(1) (2019) 841--851.

\bibitem{peng2019short}
Y.~Peng, Y.~Wang, X.~Lu, H.~Li, D.~Shi, Z.~Wang, J.~Li, Short-term load
  forecasting at different aggregation levels with predictability analysis,
  arXiv preprint arXiv:1903.10679.

\bibitem{cai2019day}
M.~Cai, M.~Pipattanasomporn, S.~Rahman, Day-ahead building-level load forecasts
  using deep learning vs. traditional time-series techniques, Applied Energy
  236 (2019) 1078--1088.

\bibitem{wang2018short}
Y.~Wang, M.~Liu, Z.~Bao, S.~Zhang, Short-term load forecasting with
  multi-source data using gated recurrent unit neural networks, Energies 11~(5)
  (2018) 1138.

\bibitem{strubell2019energy}
E.~Strubell, A.~Ganesh, A.~McCallum, Energy and policy considerations for deep
  learning in nlp, in: 57th Annual Meeting of the Association for Computational
  Linguistics (ACL), Florence, Italy, 2019.

\bibitem{schwartz2019green}
R.~Schwartz, J.~Dodge, N.~A. Smith, O.~Etzioni, Green ai, arXiv preprint
  arXiv:1907.10597.

\bibitem{dalgren2019greenml}
A.~Dalgren, Y.~Lundeg{\aa}rd, Greenml: A methodology for fair evaluation of
  machine learning algorithms with respect to resource consumption (2019).

\bibitem{dodge2019show}
J.~Dodge, S.~Gururangan, D.~Card, R.~Schwartz, N.~A. Smith, Show your work:
  Improved reporting of experimental results, in: 9th International Joint
  Conference on Natural Language Processing (IJCNLP), Hong Kong, China, 2019.

\bibitem{zhou2017pdeep}
X.-X. Zhou, W.-F. Zeng, H.~Chi, C.~Luo, C.~Liu, J.~Zhan, S.-M. He, Z.~Zhang,
  pdeep: Predicting ms/ms spectra of peptides with deep learning, Analytical
  chemistry 89~(23) (2017) 12690--12697.

\bibitem{chen2018inferring}
Y.~Chen, C.~Hu, B.~Hu, L.~Hu, H.~Yu, C.~Miao, Inferring cognitive wellness from
  motor patterns, IEEE Transactions on Knowledge and Data Engineering 30~(12)
  (2018) 2340--2353.

\bibitem{fawaz2019deep}
H.~I. Fawaz, G.~Forestier, J.~Weber, L.~Idoumghar, P.-A. Muller, Deep learning
  for time series classification: a review, Data Mining and Knowledge Discovery
  (2019) 1--47.

\bibitem{voulodimos2018deep}
A.~Voulodimos, N.~Doulamis, A.~Doulamis, E.~Protopapadakis, Deep learning for
  computer vision: A brief review, Computational intelligence and neuroscience
  2018.

\bibitem{smart2011metering}
C.~E. Smart, Metering customer behaviour trials (cbt) findings report.

\bibitem{tensorflow}
Google, Tensorflow,
  \url{https://tensorflow.google.cn/versions/r1.10/api_docs/python/tf},
  accessed June 21, 2019.

\bibitem{strubell2018linguistically}
E.~Strubell, P.~Verga, D.~Andor, D.~Weiss, A.~McCallum, Linguistically-informed
  self-attention for semantic role labeling, in: Proceedings of the 2018
  Conference on Empirical Methods in Natural Language Processing, 2018, pp.
  5027--5038.

\bibitem{campos2018skip}
V.~Campos~Camunez, B.~Jou, X.~Gir{\'o}~Nieto, J.~Torres~Vi{\~n}als, S.-F.
  Chang, Skip rnn: learning to skip state updates in recurrent neural networks,
  in: Sixth International Conference on Learning Representations: Monday April
  30-Thursday May 03, 2018, Vancouver Convention Center,
  Vancouver:[proceedings], 2018, pp. 1--17.

\bibitem{szegedy2017inception}
C.~Szegedy, S.~Ioffe, V.~Vanhoucke, A.~A. Alemi, Inception-v4, inception-resnet
  and the impact of residual connections on learning, in: Thirty-First AAAI
  Conference on Artificial Intelligence, 2017.

\bibitem{tian2018deep}
C.~Tian, J.~Ma, C.~Zhang, P.~Zhan, A deep neural network model for short-term
  load forecast based on long short-term memory network and convolutional
  neural network, Energies 11~(12) (2018) 3493.

\bibitem{kim2019recurrent}
J.~Kim, J.~Moon, E.~Hwang, P.~Kang, Recurrent inception convolution neural
  network for multi short-term load forecasting, Energy and Buildings 194
  (2019) 328--341.

\bibitem{bahdanau2014neural}
D.~Bahdanau, K.~Cho, Y.~Bengio, Neural machine translation by jointly learning
  to align and translate, arXiv preprint arXiv:1409.0473.

\bibitem{bai2018empirical}
S.~Bai, J.~Z. Kolter, V.~Koltun, An empirical evaluation of generic
  convolutional and recurrent networks for sequence modeling, arXiv preprint
  arXiv:1803.01271.

\bibitem{voss2018residential}
M.~Vo{\ss}, C.~Bender-Saebelkampf, S.~Albayrak, Residential short-term load
  forecasting using convolutional neural networks, in: 2018 IEEE International
  Conference on Communications, Control, and Computing Technologies for Smart
  Grids (SmartGridComm), IEEE, 2018, pp. 1--6.

\bibitem{he2016deep}
K.~He, X.~Zhang, S.~Ren, J.~Sun, Deep residual learning for image recognition,
  in: Proceedings of the IEEE conference on computer vision and pattern
  recognition, 2016, pp. 770--778.

\end{thebibliography}

\end{document}